\documentclass[12pt]{article}
\usepackage{eurosym}
\usepackage{graphicx}
\usepackage[utf8]{inputenc}
\usepackage[T1]{fontenc}
\usepackage{indentfirst}
\usepackage[margin=0.6in,nomarginpar]{geometry}
\usepackage[final]{hyperref}
\usepackage{amsmath}
\usepackage{hyperref}
\usepackage{cite}
\usepackage{xcolor}
\usepackage{subcaption}

\hypersetup{
colorlinks=true,
linkcolor=blue,
citecolor=blue,
filecolor=magenta,
urlcolor=blue
}

\begin{document}

\title{Van der Waals black holes in rainbow gravity}
\author{R. Oubagha \thanks{%
rabah.oubagha@univ-oeb.dz} \\
Laboratoire de syst\`{e}mes dynamiques et contr\^{o}le (L.S.D.C), \\
D\'{e}partement des sciences de la mati\`{e}re,\\
Facult\'{e} des Sciences Exactes et SNV,\\
Universit\'{e} de Oum-El-Bouaghi, 04000, Oum El Bouaghi, Algeria. \and B.
Hamil\thanks{%
hamilbilel@gmail.com(Corresponding author)} \\
Laboratoire de Physique Math\'{e}matique et Subatomique, \\
Facult\'{e}\ des Sciences Exactes, \\
Universit\'{e}\ Constantine 1, Constantine, Algeria. \and M. Merad \thanks{%
meradm@gmail.com} \\
Laboratoire de syst\`{e}mes dynamiques et contr\^{o}le (L.S.D.C), \\
D\'{e}partement des sciences de la mati\`{e}re,\\
Facult\'{e} des Sciences Exactes et SNV,\\
Universit\'{e} de Oum-El-Bouaghi, 04000, Oum El Bouaghi, Algeria.\and B. C. L\"{u}tf\"{u}o%
\u{g}lu\thanks{%
bekir.lutfuoglu@uhk.cz} \\
Department of Physics, University of Hradec Kr\'{a}lov\'{e},\\
Rokitansk\'{e}ho 62, 500 03 Hradec Kr\'{a}lov\'{e}, Czechia.}
\date{\today }
\maketitle

\begin{abstract}
{Recently, Rajagapol et al presented an asymptotically AdS black hole metric whose thermodynamics qualitatively mimics the behavior of the Van der Waals fluid by treating the cosmological constant as a thermodynamic pressure.  In some studies in the literature, authors have discussed the effects of deformed algebras such as generalized and extended uncertainty principles on the thermal quantities of these black holes. In this manuscript, we considered another deformation,  the rainbow gravity formalism,  and we investigated its impact on the Van der Waal black hole thermodynamics. To this end, we first generated the modified lapse and mass functions, and then we derived the modified thermal quantities such as thermodynamic volume, Hawking temperature, entropy, and specific heat functions. Finally, we explored the thermodynamics of a black hole, which mimics the thermodynamics of an ideal gas, under the influence of the rainbow gravity formalism.}
\end{abstract}


\newpage

\section{Introduction}

The unification of general relativity and quantum mechanics is an open problem in physics that needs to be solved. Various approaches have been used to achieve this goal, such as string theory \cite{Amati}, spacetime foam models \cite{Amelino1}, and loop quantum gravity \cite{Amelino2}. The common aspect of these models is the presence requirement of a minimum observable length, known as the Planck length, $L_{P}\sim\frac{1}{E_{P}}$, where $E_{P}$ corresponds to the Planck energy. Lorentz invariance is violated in some of these models, such as in the semi-classical limit of the loop quantum gravity \cite{Carlos}, and in the noncommutative geometry \cite{Carrol}, via a formalism based on a modified dispersion relation (MDR) \cite{1} in which the Planck energy is considered as a second relativistic invariant in addition to the speed of the light, $c$. Taking this change into account, the special theory of relativity can be considered as the classical limit of a more general formalism, namely doubly special relativity (DSR) \cite{3}. This approach was proven to be effective for the explanation of anomalies observed in ultra-high-energy cosmic rays \cite{Mague} and TeV photons\cite{2}. 

According to Magueijo and Smolin, the gravitational force should exhibit different behaviors or colors depending on the energy of the particles involved. From this perspective, they proposed a generalization of DSR and introduced the concept of Rainbow Gravity ({RaGr}) to describe gravity in curved space-time \cite{4}. Their main purpose in this framework was to preserve the relativity of inertial frames while preserving the Planck energy as a universal constant for all inertial observers. This assumption is achieved through the implementation of a non-linear Lorentz transformation in momentum space, which gives rise to the following form of the MDR
\begin{eqnarray}
E^{2}\mathcal{F}\left( \frac{E%
}{E_{P}}\right) ^{2}+p^{2}\mathcal{G}\left( \frac{E}{E_{P}}\right) ^{2}=m^{2},    \end{eqnarray}
for a test particle of mass, $m$, and energy, $E$, with $c=1$. Here, $\mathcal{F}\left( \frac{E}{E_{P}}\right) $ and $\mathcal{G}\left(\frac{E}{E_{P}}\right) $ are known as rainbow functions \cite{1,2,3,4,Mague}. At this point, we should emphasize that the rainbow functions tend to be equal to one in the infrared limit, and hence, the MDR relation takes to the standard form of the energy dispersion relation. On the other hand, the restriction imposed by the Planck energy makes it impossible for the test particle to reach energies beyond this limit. The reason behind the limitation lies in the differential effects of gravity on particles with varying energies. This perspective brings a new interpretation to the definition of spacetime: relativistic locality. From a mathematical viewpoint, the Planck energy, which can be seen as the curvature of the momentum space, requires the description of spacetime via Finsler geometry \cite{7}. In the description of this formalism through non-commutative geometry, the symmetry groups must be replaced by their non-commutative analogs, the Hopf algebras\cite{6}.

Black hole solutions are among the most interesting predictions of general relativity. Historically, Karl Schwarzschild was the first scientist, 
originally an astronomer,  to predict their existence by solving Einstein's equations for the vacuum \cite{Karl1, Karl2}. He introduced the black hole concept by characterizing a spacetime region from which nothing can run away, known as the event horizon. Essentially, he thought that the singularity located at the origin of the radial coordinate of the region makes it possible to enter that region but impossible to escape. Since black holes were believed to be non-observable, all black hole solutions obtained after that time, including the Schwarzschild black hole, were often stated as a mystery in research papers. But this belief began to change fifty years ago with the innovative contributions of Hawking and Bekenstein in relating the laws of black hole physics to the ordinary laws of thermodynamics \cite{Bekenstein1, Bekenstein2, Hawking, Bek, Hawking1, Hawk}. Following the pioneering idea that black holes can have both temperature and entropy, examining the thermodynamic properties of black holes has become an important field of research \cite{Aman2003, Dolan2011, Carlip2014, Mann2015, Kumar2020}, and this fact enabled a more in-depth discussion about quantum gravity and its effect \cite{0012, Nouicer2007, Myung2007, Gangopadhyay2014, 18, 19, cc14, cc16,  cc19, cc20, cc21, cc22, cc23, cc25, Marcos1, Marcos2, Marcos3, erk1, erk2, erk3, erk4, erk5, erk6, erk7}.

The study of asymptotically anti-de Sitter (AdS) black holes has collected significant attention, mainly due to the AdS/CFT correspondence \cite{Ray1, Ray2, Ray3}. A fundamental characteristic of this type of black hole is the recognition of a negative cosmological constant, $\Lambda $, that results in a positive thermodynamic pressure, $P$, \cite{Teitelboim}.  The extended black hole thermodynamics framework has brought forth numerous intriguing additions to the study of thermodynamic phenomena in AdS black holes \cite{Hubeny}. When the AdS black hole incorporates charge and/or rotation, its characteristics exhibit a qualitative resemblance to that of the Van der Waals (VdW) fluid \cite{Chamblin, Cham}. In Refs. \cite{Rajagopal, Hendi}, the authors have observed a first-order phase transition reminiscent of VdW behavior. This particular type of phase transition becomes more evident in the extended phase space, where the cosmological constant is interpreted as a thermodynamic pressure. From the VDW black hole models, extremely interesting formalisms have emerged such as anyonic black holes \cite{Abc}. Recently, Ökcü and Aydiner investigated the thermodynamics of VdW black holes by handling the Generalized Uncertainty Principle instead of the conventional Heisenberg formalism \cite{Aydiner}. Inspired by their work, we examined the thermodynamics of VdW black holes by employing the Extended Uncertainty Principle (EUP) formalism \cite{OurVdW}. Remembering the distinct features of the {RaGr} formalism compared to the GUP and EUP formalisms {\cite{Li2017, Junior2020, B1, B2, B3, B4, B5, Gim2014, Ditta2023,ref21,ref22}}, in this manuscript we aim to study the VdW black hole thermodynamics in the context of the {RaGr}. To this end, we construct the manuscript as follows: In Section \ref{sec2}, we provide a concise introduction to the {RaGr} formalism and examine its impact on the Hawking temperature, employing a semi-classical approach. Next, in Section \ref{sec3}, we obtained the {RaGr}-corrected thermodynamics of VdW black holes. Then, in Section  \ref{sec4}, with a special parameter choice we examine the {RaGr} effects on the thermodynamics of a black hole whose thermodynamics matches with the thermodynamics of an ideal gas. Finally, in the last Section, we conclude the manuscript with a summary of our findings.

\section{RaGr-corrected Hawking temperature}\label{sec2}

In this section, we want to briefly show how one can give a generic {RaGr} correction to the semi-classical Hawking temperature. To this end, we start by considering a static spherically symmetric black hole metric
\begin{equation}
ds^{2}=-f\left( r\right) dt^{2}+\frac{1}{f\left( r\right) }%
dr^{2}+r^{2}d\Omega ^{2},  \label{metric}
\end{equation}%
where $f(r)$ and $d\Omega ^{2}=d\theta ^{2}+\sin ^{2}\theta d\phi ^{2}$ are the lapse function and the line element of two-dimensional hypersurfaces, respectively. To introduce the {RaGr} effects, we adhere to the approach outlined in \cite{1}, and modify the metric via two arbitrary rainbow functions as $dt\rightarrow \frac{dt}{\mathcal{F}\left( \frac{E}{E_{P}}\right) }$ and $dx_{i}\rightarrow \frac{dx_{i}}{\mathcal{G}\left( \frac{E}{E_{P}}\right) }$. In so doing, we arrive at the {RaGr}-modified metric of the following form:
\begin{equation}
ds^{2}=-\frac{f\left( r\right) }{\mathcal{F}^{2}\left( \frac{E}{E_{P}}%
\right) }dt^{2}+\frac{1}{\mathcal{G}^{2}\left( \frac{E}{E_{P}}\right)
f\left( r\right) }dr^{2}+\frac{r^{2}}{\mathcal{G}^{2}\left( \frac{E}{E_{P}}%
\right) }d\Omega ^{2},  \label{RM}
\end{equation}
where the  rainbow functions obey the following relationship
\begin{equation}
\lim_{\frac{E}{E_{P}}\rightarrow 0}\mathcal{F}\left( \frac{E}{E_{P}}\right)
=\lim_{\frac{E}{E_{P}}\rightarrow 0}\mathcal{G}\left( \frac{E}{E_{P}}\right)
=1.
\end{equation}%
Based on Eq. (\ref{RM}), we can derive the {RaGr}-corrected Hawking temperature from the definition of surface gravity 
\begin{equation}
T_{H}=\frac{\kappa }{4\pi }=\frac{\kappa _{0}}{4\pi }\frac{\mathcal{G}\left( 
\frac{E}{E_{P}}\right) }{\mathcal{F}\left( \frac{E}{E_{P}}\right) },
\label{RT}
\end{equation}%
where $\kappa _{0}=\left. \frac{df\left( r\right) }{dr}\right\vert _{r=r_{H}} $ is the standard surface gravity. 

Now, we want to eliminate the energy dependence of the probe particle. According to \cite{1}, the conventional Heisenberg's uncertainty principle (HUP), $\Delta p\geq \frac{1}{\Delta x}$,  also holds in the {RaGr} context. Consequently, the HUP can be transformed into a minimum limit on the energy of a particle released through the Hawking radiation, $E\geq \frac{1}{\Delta x}$. Then,
near the black hole horizon, we can take the positional uncertainty to be equivalent to the radius of the black hole, $\Delta x\simeq r_{H}$.
Therefore, we get 
\begin{equation}
E\simeq \frac{1}{r_{H}}.  \label{HR}
\end{equation}%
Then, by substituting Eq. (\ref{HR}) into Eq. (\ref{RT}), we express the {RaGr}-corrected Hawking temperature in the form of
\begin{equation}
T_{H}=\frac{\kappa }{4\pi }=\frac{f^{\prime }\left( r_{H}\right) }{4\pi }%
\frac{\mathcal{G}\left( \frac{1}{r_{H}E_{P}}\right) }{\mathcal{F}\left( 
\frac{1}{r_{H}E_{P}}\right) }.
\end{equation}

\section{RaGr-corrected thermodynamics of VdW black holes}\label{sec3}

In this section, we explore the impacts of the {RaGr} formalism on the thermodynamics of  VdW black holes. To achieve this goal, we liken the VdW black hole to a VdW fluid, so, we start with the equation of the state of a VdW fluid, 
\begin{equation}
T=\left( P+\frac{a}{v^{2}}\right) \left( v-b\right) .  \label{eq}
\end{equation}
Here, $T$ and $P$ represent the temperature and the pressure, and $v$ is the specific volume which can be written in terms of the volume, $V$, and the degrees of freedom, $N$, as $v=6V/N$. The other two parameters, namely $a$ and $b$, are two positive constants that stand for the degree of attraction existing between particles and their volume, respectively. 

Now, we have to determine the {RaGr}-corrected lapse function of the VdW black holes. In the conventional case, which corresponds to Eq. \eqref{metric}, we could consider the lapse function with the following form  
\begin{equation}
f\left( r\right) =\frac{r^{2}}{l^{2}}-\frac{2M}{r}-h\left( r,P\right) .
\label{f}
\end{equation}
where the thermodynamic pressure could be expressed in terms of the negative cosmological constant
\begin{equation}
P=\frac{3}{8\pi l^{2}}=-\frac{\Lambda }{8\pi },
\end{equation}
and the given metric should be the solution of Einstein's field equations defined by
\begin{equation}
G_{\mu \nu }+\Lambda g_{\mu \nu }=8\pi T_{\mu \nu }.  \label{Einstein}
\end{equation}
Therefore, in the presence of {RaGr} formalism the components of the Ricci tensor, $R_{\mu\nu}$, must be deduced from the metric given in Eq. (\ref{RM}). In this case, we get 
\begin{eqnarray}
R_{00}&=&\frac{\mathcal{G}^{2}\left( \frac{E}{E_{P}}\right) }{\mathcal{F}%
^{2}\left( \frac{E}{E_{P}}\right) }\frac{f}{2r}\left[ rf\text{ }^{\prime
\prime }+2f\text{ }^{\prime }\right] , \\
R_{11}&=&-\frac{rf\text{ }^{\prime \prime }+2f\text{ }^{\prime }}{2rf}, \\
R_{22}&=&1-rf\text{ }^{\prime }-f, \\
R_{33}&=&\sin ^{2}\theta \left[ 1-rf^{\prime }-f\right] ,
\end{eqnarray}
and the associated Ricci scalar, $R$, reads:
\begin{equation}
R=\mathcal{G}^{2}\left( \frac{E}{E_{P}}\right) \frac{2-r^{2}f\text{ }%
^{\prime \prime }-4rf\text{ }^{\prime }-2f}{r^{2}}.
\end{equation}
Then, from Eq. (\ref{Einstein}) we get the stress-energy tensor components 
\begin{equation}
\rho =-p_{1}=\mathcal{G}^{2}\left( \frac{E}{E_{P}}\right) \frac{1-rf\text{ }%
^{\prime }-f}{8\pi r^{2}}+P .
\end{equation}%
\begin{equation}
p_{2}=p_{3}=\mathcal{G}^{2}\left( \frac{E}{E_{P}}\right) \frac{2f\text{ }%
\prime +rf\text{ }"}{16\pi r}-P .
\end{equation}
Here, we choose the stress-energy tensor as an anisotropic fluid source in the following form
\begin{equation}
T_{\mu \nu }=\rho e_{\mu }^{0}e_{\nu }^{0}+\sum_{i}p_{i}e_{\mu }^{i}e_{\nu
}^{i},
\end{equation}
where $e_{\mu i}$, $\rho $, and $p_{i}$ represent the elements of the vielbein $(i=1,2,3)$, the energy density, and the principal pressure, respectively.

\begin{figure}[tbh!]
\begin{minipage}[t]{0.5\textwidth}
        \centering
        \includegraphics[width=\textwidth]{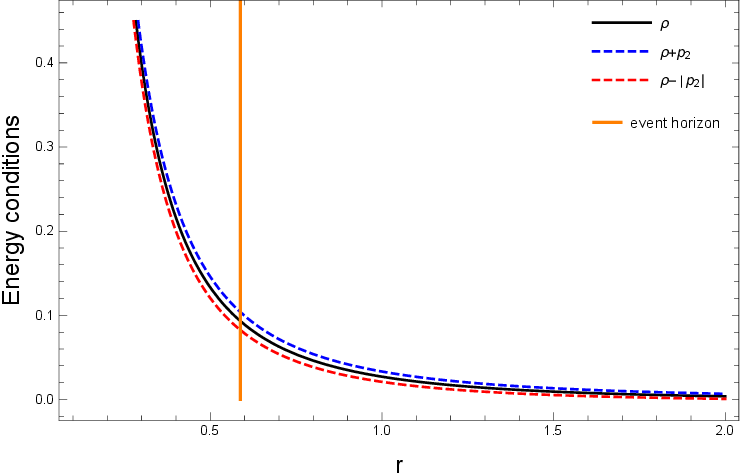}
       \subcaption{$P=0.1$.} \label{figcona}
   \end{minipage}%
\begin{minipage}[t]{0.50\textwidth}
        \centering
        \includegraphics[width=\textwidth]{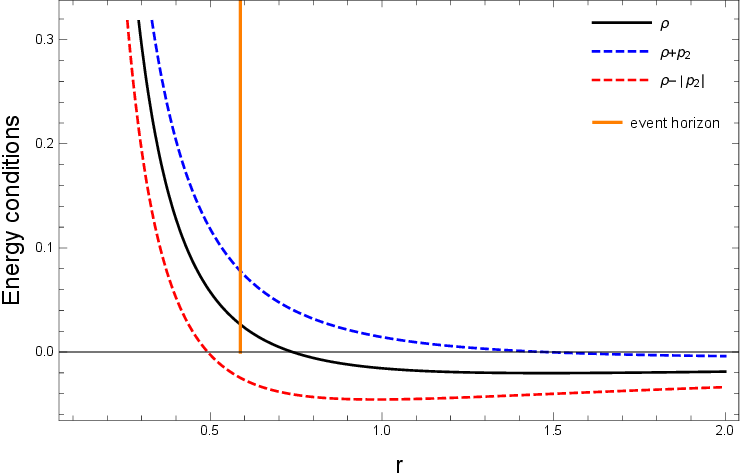}\\
        \subcaption{ $P=0.5$. }\label{figconb}
    \end{minipage}\hfill
\caption{The energy conditions for $a=0.01$, $b=\frac{3}{8\pi }$, $M=0.1$, $\gamma =\epsilon=0.05$, $E_{p}=1$.}
\label{con}
\end{figure}

On the other side, it is a well-known fact that the mass of a black hole can be determined through, $f\left( r_{H}\right) =0$. Thus, in our case, we get
\begin{equation}
M=\frac{4}{3}\pi r_{H}^{3}P-\frac{r_{H}}{2}h\left( r_{H},P\right) .
\label{M}
\end{equation}
Then, we can express the thermodynamic volume by employing the following thermodynamic relation:
\begin{equation}
V=\frac{\partial M}{\partial P}. \label{VM}
\end{equation}
We find%
\begin{equation}
V=\frac{\partial M}{\partial P}=\frac{4}{3}\pi r_{H}^{3}-\frac{r_{H}}{2}%
\frac{\partial }{\partial P}h\left( r_{H},P\right) ,  \label{V}
\end{equation}
so that, the specific volume reads:
\begin{equation}
v=\frac{6}{4\pi r_{H}^{2}}\left[ \frac{4}{3}\pi r_{H}^{3}-\frac{r_{H}}{2}%
\frac{\partial }{\partial P}h\left( r_{H},P\right) \right] .  \label{SV}
\end{equation}%
Then, by substituting Eq. (\ref{SV}) in Eq. (\ref{eq}), we obtain {RaGr}-corrected Hawking temperature of the VdW black hole in the form of: 
\begin{equation}
T_{H}=\frac{1}{4\pi }\left( 8\pi r_{H}P-\frac{1}{r_{H}}h\left( r_{H},P\right) -%
\frac{\partial }{\partial r_{H}}h\left( r_{H},P\right) \right) \frac{%
\mathcal{G}\left( \frac{1}{r_{H}E_{P}}\right) }{\mathcal{F}\left( \frac{1}{%
r_{H}E_{P}}\right) }.  \label{Tem}
\end{equation}
{Now, we need to determine the indefinite function, $h\left( r, P\right),$ that establishes the connection between the black hole thermodynamics and the VdW fluid. Here, it is worth emphasizing that the uncertain function has to be compatible with the lapse and equation of state functions. In \cite{Rajagopal}, Rajagopal et al. proposed a polynomial form for the solution of the indefinite function. In particular, they used a first-order polynomial with respect to $P$, since other forms constructed with higher powers of $P$ terms violate the asymptotic AdS structure \cite{Delsate2015}. Here, we follow their ansatz and consider}
\begin{equation}
h(r,P)=A(r)-PB(r),  \label{h}
\end{equation}%
with two arbitrary functions $A(r)$ and $B(r)$. Then, we substitute Eq. (\ref{h}) in Eq. (\ref{Tem}), and we find
\small
\begin{eqnarray}
 T=\frac{1}{4\pi }\left( -\frac{1}{r_{H}}A\left( r_{H}\right) -\frac{\partial 
}{\partial r_{H}}A\left( r_{H}\right) \right) \frac{\mathcal{G}\left( \frac{1%
}{r_{H}E_{P}}\right) }{\mathcal{F}\left( \frac{1}{r_{H}E_{P}}\right) }+\frac{%
P}{4\pi }\left( 8\pi r_{H}+\frac{1}{r_{H}}B\left( r_{H}\right) +\frac{%
\partial }{\partial r_{H}}B\left( r_{H}\right) \right) \frac{\mathcal{G}%
\left( \frac{1}{r_{H}E_{P}}\right) }{\mathcal{F}\left( \frac{1}{r_{H}E_{P}}%
\right) }.  \label{htem}
\end{eqnarray}
\normalsize
Then, by comparing the temperature expression in Eqs. (\ref{eq}) and (\ref{htem}), we write
\begin{equation}
PF_{1}\left( r_{H}\right) +F_{2}\left( r_{H}\right) =0, \label{Wag}
\end{equation}
where
\begin{equation}
F_{1}\left( r_{H}\right) =\left( 2r_{H}+\frac{1}{4\pi r_{H}}B\left(
r_{H}\right) +\frac{1}{4\pi }\frac{\partial }{\partial r_{H}}B\left(
r_{H}\right) \right) \frac{\mathcal{G}\left( \frac{1}{r_{H}E_{P}}\right) }{%
\mathcal{F}\left( \frac{1}{r_{H}E_{P}}\right) }-\left(2r_{H}+\frac{3}{4\pi r_{H}}%
B\left( r_{H}\right) -b\right),  \label{fun}
\end{equation}%
and%
\begin{equation}
F_{2}\left( r_{H}\right) =\frac{1}{4\pi }\left( \frac{1}{r_{H}}A\left(
r_{H}\right) +\frac{\partial }{\partial r_{H}}A\left( r_{H}\right) \right) 
\frac{\mathcal{G}\left( \frac{1}{r_{H}E_{P}}\right) }{\mathcal{F}\left( 
\frac{1}{r_{H}E_{P}}\right) }+\left( v-b\right) \frac{a}{v^{2}}.
\end{equation}
Here, we notice that Eq. \eqref{Wag} could be satisfied if these two functions vanish simultaneously. On the other hand, 
this two-ordinary differential equation system depends on the choice of the {RaGr} functions. Therefore to go further, we have to consider particular {RaGr} functions. In this manuscript, we choose the following ones that possess a physical background in string theory, loop quantum gravity, and quantum cosmology  \cite{Amelino1, Amelino2}: 
\begin{eqnarray}
\mathcal{G}\left( \frac{E}{E_{P}}\right) =\sqrt{1-\gamma \frac{E^{2}}{
E_{P}^{2}}}; \qquad
\mathcal{F}\left( \frac{E}{E_{P}}\right) =1.  \label{rai}
\end{eqnarray}
By substituting Eq. (\ref{rai}) into Eq. (\ref{fun}), we obtain
\begin{equation}
B\left( r_{H}\right) =-\frac{8}{3}\pi r_{H}^{2}+\frac{4\pi br_{H}}{3}+\frac{%
8\pi }{9}br_{H}^{2}\left( \frac{3}{r_{H}}-\frac{3\gamma }{2E_{P}^{2}r_{H}^{3}%
}\right) +c_{0}r_{H}^{2}\left( 1-\frac{3\gamma }{4E_{P}^{2}r_{H}^{2}}\right)
+\mathcal{O}\left( \gamma ^{2}\right) ,  \label{bfun}
\end{equation}
where $c_{0}$ is a constant of integration. To determine it, we go to the limit of $\gamma =0$, where Eq. (\ref{bfun}) reduces to the standard result given in \cite{Rajagopal}.  In so doing, we find $c_{0}=\frac{8\pi }{3}$. Then, Eq. (\ref{bfun}) reads: 
\begin{equation}
B\left( r_{H}\right) \simeq 4b\pi r_{H}-\frac{2\pi \gamma }{E_{P}^{2}}\left(
1+\frac{2b}{3r_{H}}\right) .  \label{B}
\end{equation}
Next, we solve the second equation, $F_{2}=0$.  After following simple algebra,  we obtain
\begin{equation}
\frac{\partial }{\partial r_{H}}A\left( r_{H}\right) =\frac{4\pi a\left(
b-2r_{H}-\frac{3}{4\pi r_{H}}B\left( r_{H}\right) \right) }{\left( 2r_{H}+%
\frac{3}{4\pi r_{H}}B\left( r_{H}\right) \right) ^{2}\sqrt{1-\gamma \frac{1}{%
r_{H}^{2}E_{P}^{2}}}}-\frac{1}{r_{H}}A\left( r_{H}\right) ,
\end{equation}
with its solution 
\begin{eqnarray}
A\left( r_{H}\right) &=&-2\pi a+\frac{3\pi ab^{2}}{r_{H}\left(
2r_{H}+3b\right) }+\frac{29}{9}\frac{\pi a\gamma }{E_{P}^{2}\left(
2r_{H}+3b\right) }+\frac{4\pi ab}{r_{H}}\ln \left( \frac{3}{2}+\frac{r_{H}}{b%
}\right) +\frac{16}{27}\frac{\pi a\gamma }{bE_{P}^{2}r_{H}}\ln \left(
2r_{H}+3b\right)  \notag \\
&&-\frac{16}{27}\frac{\pi a\gamma }{bE_{P}^{2}r_{H}}\ln r-\frac{5}{3}\frac{%
\pi ab\gamma }{E_{P}^{2}r_{H}\left( 2r_{H}+3b\right) ^{2}}+\mathcal{O}\left( \gamma ^{2}\right).  \label{A}
\end{eqnarray}
We take the solutions of $A$ and $B$ up to the first order of $\gamma$ and substitute them in Eq. (\ref{h}). In so doing, we arrive at the expression of $h\left( r,P\right) $ in the following form:
\begin{eqnarray}
h(r,P) &=&-2\pi a+\frac{3\pi ab^{2}}{r\left( 2r+3b\right) }+\frac{29}{9}%
\frac{\pi a\gamma }{E_{P}^{2}\left( 2r+3b\right) }+\frac{4\pi ab}{r}\ln
\left( \frac{3}{2}+\frac{r}{b}\right) +\frac{16}{27}\frac{\pi a\gamma }{%
bE_{P}^{2}r}\ln \left( 2r+3b\right)  \notag \\
&&-\frac{5}{3}\frac{\pi ab\gamma }{E_{P}^{2}r\left( 2r+3b\right) ^{2}}-\frac{%
16}{27}\frac{\pi a\gamma }{bE_{P}^{2}r}\ln r-4Pb\pi r+\frac{2P\gamma \pi }{%
E_{P}^{2}}\left( 1+\frac{2b}{3r}\right).  \label{hrp}
\end{eqnarray}
Finally, by incorporating the expression of $h(r,P)$ into Eq. (\ref{f}), we get the {RaGr}-corrected lapse function in the form of 
\begin{eqnarray}
f\left( r\right) &=&\frac{8\pi P}{3}r^{2}-\frac{2M}{r}+2\pi a-\frac{3\pi
ab^{2}}{r\left( 2r+3b\right) }-\frac{29}{9}\frac{\pi a\gamma }{%
E_{P}^{2}\left( 2r+3b\right) }-\frac{4\pi ab}{r}\ln \left( \frac{3}{2}+\frac{%
r}{b}\right)  \notag \\
&-&\frac{16}{27}\frac{\pi a\gamma }{bE_{P}^{2}r}\ln \left( 2r+3b\right) +%
\frac{16}{27}\frac{\pi a\gamma }{bE_{P}^{2}r}\ln r+\frac{5}{3}\frac{\pi
ab\gamma }{E_{P}^{2}r\left( 2r+3b\right) ^{2}} +4Pb\pi r-\frac{2P\gamma \pi }{E_{P}^{2}}\left( 1+\frac{2b}{3r}\right) . \label{RGlapse}
\end{eqnarray}
{
Here, we presume that the considered metric is a solution of the Einstein field equations with a given energy--momentum source. For the energy-momentum source to be physically meaningful, we need it to satisfy particular energy conditions such as positivity of energy density and dominance of the energy density over the pressure. In \cite{Rajagopal}, the authors noted that the minimal requirement is the weak energy condition. Now, let us express these energy conditions:
\begin{equation}
\text{Weak: \ \ \ \ \ \ \ \ \ }\rho \geq 0;\text{ \ }\rho +p_{i}\geq 0,
\end{equation}%
\begin{equation}
\text{Strong: \ \ }\rho +p_{i}\geq 0;\text{ \ }\rho
+\sum_{i}p_{i}\geq 0,
\end{equation}%
\begin{equation}
\text{Dominant: \ \ \ \ \ \ }\rho \geq 0;\text{ \ }\rho \geq \left\vert
p_{i}\right\vert ,
\end{equation}
and examine them with the help of Fig. \ref{con} . }

{
In Fig. \ref{figcona}, where $P=0.1$, we observe that all the energy conditions may be satisfied simultaneously near the horizon. However, Fig. \ref{figconb} shows that as the pressure increases to $P=0.5$, $\rho $ diminishes at small radii, resulting in a violation of the dominant energy condition. It is worth emphasizing that RaGr corrections do not alter the weak energy condition, as in the non-corrected case reported in \cite{Rajagopal}. Therefore, we conclude that our solution meets the minimal requirement, hence, it is physically valid near the horizon.}

\begin{figure}[tbh!]
\begin{minipage}[t]{0.5\textwidth}
        \centering
        \includegraphics[width=\textwidth]{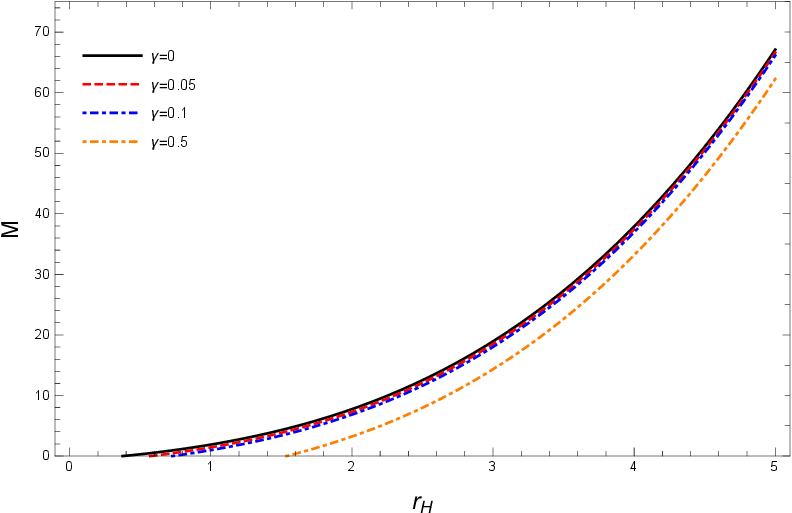}
       \subcaption{$ a=1$, $b=3/8\pi$.} \label{figRMassa}
   \end{minipage}%
\begin{minipage}[t]{0.50\textwidth}
        \centering
        \includegraphics[width=\textwidth]{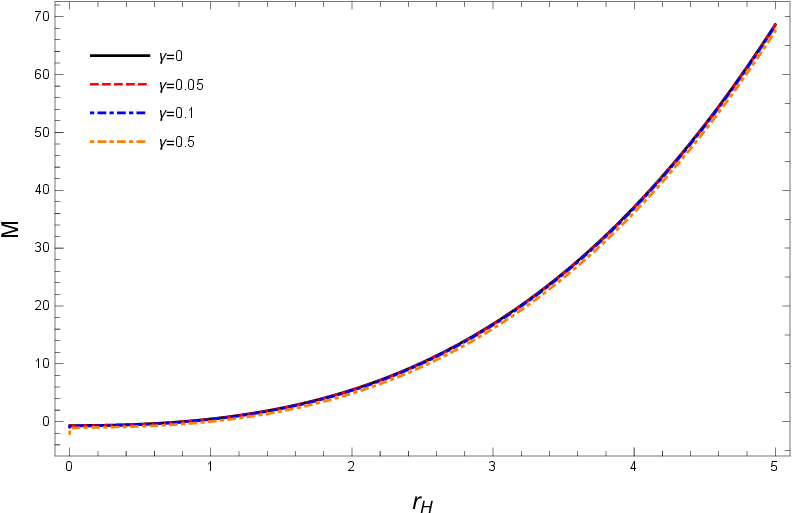}\\
        \subcaption{ $ a=1/2\pi$, $b=1$. }\label{figRMassb}
    \end{minipage}\hfill
\caption{The variation of {RaGr}-corrected mass function versus horizon for $P=0.1.$}
\label{figRMass}
\end{figure}

Now, we employ Eq. (\ref{hrp}) in Eq.(\ref{M}) and express the {RaGr}-corrected mass of a VdW black hole in the following form:
\small
\begin{eqnarray}
M &=&\frac{4}{3}\pi P r_{H}^{3}+\pi ar_{H}-\frac{3}{2}\frac{\pi ab^{2}}{%
\left( 2r_{H}+3b\right) }-\frac{29}{18}\frac{\pi a\gamma r_{H}}{%
E_{P}^{2}\left( 2r_{H}+3b\right) }-2\pi ab\ln \left( \frac{3}{2}+\frac{r_{H}%
}{b}\right) -\frac{8}{27}\frac{\pi a\gamma }{bE_{P}^{2}}\ln \text{ }\left(
2r_{H}+3b\right)  \notag \\
&&+\frac{8}{27}\frac{\pi a\gamma }{bE_{P}^{2}}\ln r_{H}+\frac{5}{6}\frac{\pi
ab\gamma }{E_{P}^{2}\left( 2r_{H}+3b\right) ^{2}}+2\pi P b r_{H}^{2}-\frac{\pi
r_{H}P\gamma }{E_{P}^{2}}\left( 1+\frac{2b}{3r_{H}}\right) .  \label{mass}
\end{eqnarray}\normalsize
It is worth noting that for $\gamma=0$ the {RaGr} modification disappears. In this case, we get
\begin{eqnarray}
M &=&\frac{4}{3}\pi P r_{H}^{3}+\pi ar_{H}-\frac{3}{2}\frac{\pi ab^{2}}{%
\left( 2r_{H}+3b\right) }-2\pi ab\ln \left( \frac{3}{2}+\frac{r_{H}%
}{b}\right) +2\pi Pb r_{H}^{2}. \label{normalmass}
\end{eqnarray}
This result is the same as the result given in Eq.  25 in \cite{OurVdW} if the EUP correction terms are omitted. Before moving on to other thermal quantities, in Fig. \ref{figRMass}, we depict the semiclassical and {RaGr}-corrected mass versus horizon to visualize the impact of the {RaGr} formalism.  

We observe that in both cases the mass function increases monotonically. However, in Fig. \ref{figRMassa}, we see that the {RaGr}-corrected mass functions always take lower values than the semi-classical mass functions. { We observe that this result is the same as the result obtained in other studies in the literature. \cite{Aydiner, Ditta2023, OurVdW}.} We also see that the {RaGr} formalism leads to a greater valued constraint on the horizon radius. In another figuration, as shown in  Fig. \ref{figRMassb}, we find that the {RaGr} effects can be negligible. 

Next, we use Eq. \eqref{VM} to derive the {RaGr}-corrected thermodynamic volume. We find
\begin{equation}
V=\frac{4}{3}\pi r_{H}^{3}+2b\pi r_{H}^{2}-\frac{r_{H}\gamma \pi }{E_{P}^{2}}%
\left( 1+\frac{2b}{3r_{H}}\right) .
\end{equation}
Note that the {RaGr} correction applied to the thermodynamic volume counts for a negative value. As a result of the {RaGr} formalism, the volume remains smaller than its semiclassical analog {similar to the EUP scenario \cite{OurVdW}. This contrasts with the outcomes from thermodynamic corrections predicted by the GUP formalism, where it was established that the corrections consistently induce the volume  \cite{Aydiner}.} 

\begin{figure}[tbh!]
\begin{minipage}[t]{0.5\textwidth}
        \centering
        \includegraphics[width=\textwidth]{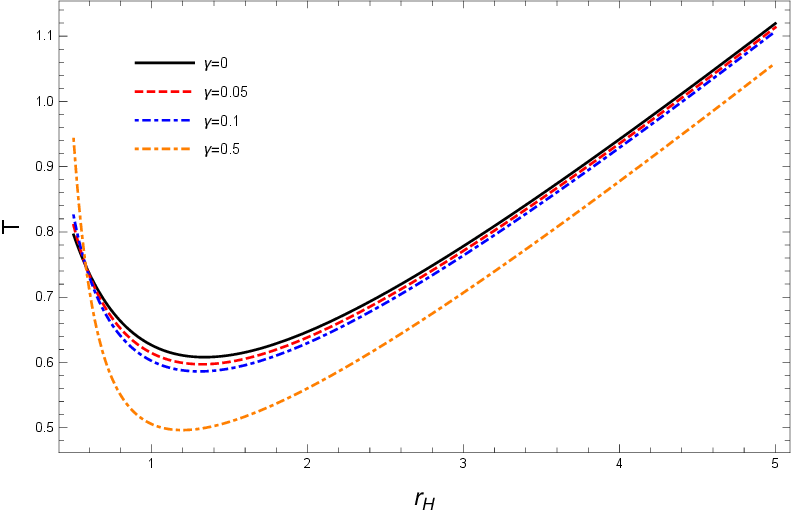}
       \subcaption{ $ a=1$, $b=3/8\pi$.}\label{figRTa}
   \end{minipage}%
\begin{minipage}[t]{0.5\textwidth}
        \centering
        \includegraphics[width=\textwidth]{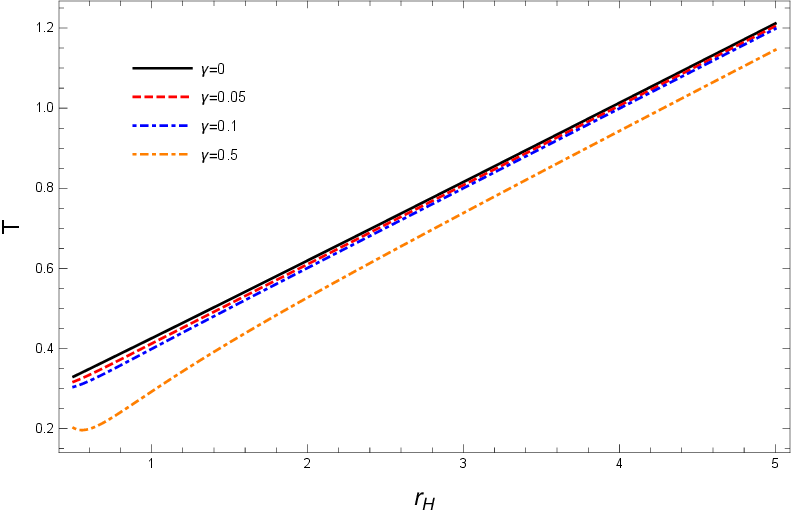}\\
        \subcaption{ $ a=1/2\pi$, $b=1$. }\label{figRTb}
    \end{minipage}\hfill
\caption{The variation of the {RaGr}-corrected Hawking temperature versus $r_{H}$
for $P=0.1.$}
\label{figrainbowT}
\end{figure}

Now, by using Eqs. (\ref{B}) and (\ref{A}), we express the {RaGr}-corrected temperature function.
\begin{eqnarray}
T &=&\left( 2bP+2r_{H}P+\frac{a}{2r_{H}}\right) \left( 1-\frac{\gamma }{%
2r_{H}^{2} {E_{P}^{2}}}\right) +\frac{4a\gamma }{27br^{2}_{H}{E_{P}^{2}}}-\frac{P\gamma }{2r_{H}{E_{P}^{2}}} 
\notag \\
&&-\frac{a}{4br_{H}(3b+2r_{H})}\left( \frac{32\gamma }{27{E_{P}^{2}}}+\frac{29 \gamma b}{9{E_{P}^{2}}}%
+8b^{2}\left( 1-\frac{\gamma }{2r_{H}^{2}{E_{P}^{2}}}\right) \right)  \notag \\
&&+\frac{a}{4(3b+2r_{H})^{2}}\left( \frac{58\gamma }{9{E_{P}^{2}}}+\frac{6b^{2}}{r_{H}}%
\left( 1-\frac{\gamma }{2r_{H}^{2}{E_{P}^{2}}}\right) \right) -\frac{5ab\gamma }{%
3r(3b+2r_{H})^{3}}.  \label{tem}
\end{eqnarray}
To present a comparison between the semi-classical and {RaGr}-corrected cases, we depict Fig. \ref{figrainbowT}.

We see that in  Fig. \ref{figRTa} the temperature first decreases, and then, increases. However, in Fig. \ref{figRTb} the temperature monotonically increases. This characteristic behavior does not alter in the presence of {RaGr} formalism. On the other hand, the {RaGr}-corrected temperature functions take smaller values compared to the semi-classic ones. { We observe that even if the temperature decreases in a certain range in both cases, it always takes positive values, which implies that a physical solution is possible for all temperatures. When we look at other studies in the literature, we notice that this finding is valid in \cite{Ditta2023}, but not in \cite{Aydiner, OurVdW}.}

Next, we drive the {RaGr}-corrected entropy function by utilizing the conventional definition.
\begin{equation}
S=\int \frac{dM}{T}.
\end{equation}
Using Eqs. (\ref{mass}) and (\ref{tem}), we simply derive the {RaGr}-corrected entropy function in the following form.
\begin{equation}
S\simeq \pi r_{H}^{2}+\frac{\gamma }{E_{P}^{2}}\ln r_{H}+\mathcal{O}\left(
\gamma ^{2}\right) .
\end{equation}
We note that the second term generates an increase in entropy, which differs from the results obtained via the GUP \cite{Aydiner} and the EUP corrected \cite{OurVdW} formalism. 

\begin{figure}[tbh]
\begin{minipage}[t]{0.5\textwidth}
        \centering
        \includegraphics[width=\textwidth]{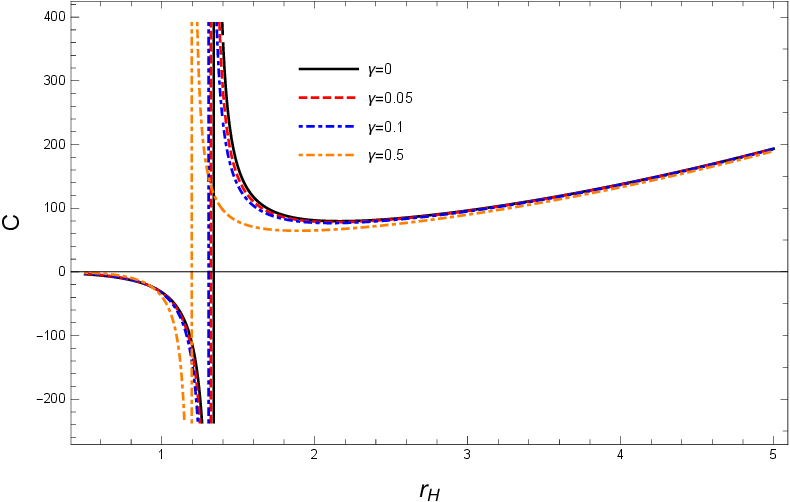}
       \subcaption{$ a=1$, $b=3/8\pi$.}\label{figRCa}
   \end{minipage}%
\begin{minipage}[t]{0.50\textwidth}
        \centering
        \includegraphics[width=\textwidth]{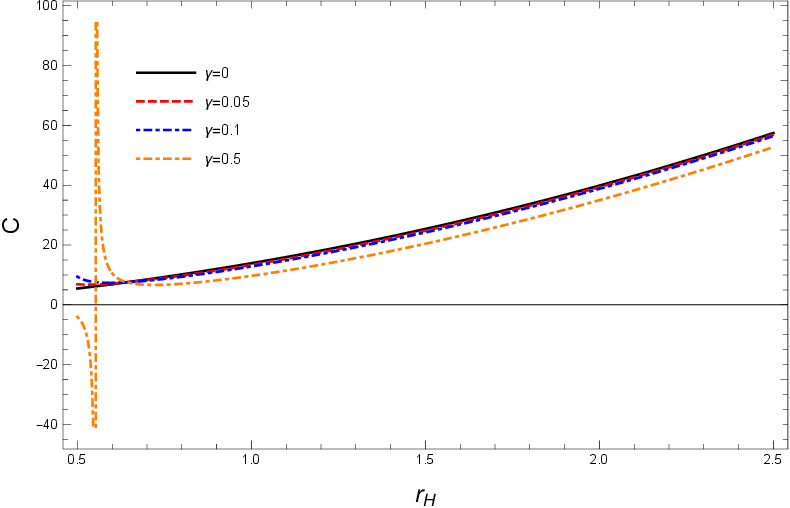}\\
        \subcaption{$ a=1/2\pi$, $b=1$.}\label{figRCb}
    \end{minipage}\hfill
\caption{The variation of the {RaGr}-corrected heat capacity versus $r_{H}$ for $P=0.1.$ }
\label{figRC}
\end{figure}

In this section finally, we investigate the {RaGr}-corrected heat capacity function to discuss the black hole stability and possible phase transitions \cite{Ma}. To this end, first, we derive the function by employing the following thermodynamic relation.
\begin{equation}
C=\frac{dM}{dT}.
\end{equation}
We find it in the closed form
\begin{equation}
C=\frac{4\pi \sqrt{1-\frac{\gamma }{r_{H}^{2}E_{P}^{2}}}\left( 4\pi
r_{H}^{2}P-\frac{1}{2}h\left( r_{H},P\right) -\frac{r_{H}}{2}\frac{\partial 
}{\partial r_{H}}h\left( r_{H},P\right) \right) }{\left( 8\pi P+\frac{1}{%
r_{H}^{2}}h\left( r_{H},P\right) -\frac{1}{r_{H}}\frac{\partial }{\partial
r_{H}}h\left( r_{H},P\right) -\frac{\partial ^{2}}{\partial r_{H}^{2}}%
h\left( r_{H},P\right) \right) +\frac{\gamma }{r_{H}^{2}E_{P}^{2}}\left( 
\frac{\partial ^{2}}{\partial r_{H}^{2}}h\left( r_{H},P\right) -\frac{2}{%
r_{H}^{2}}h\left( r_{H},P\right) \right) } .
\end{equation}
The derived form is very complex. Therefore, we need to examine it with numerical methods. To this end, we provide a visual representation in Figure (\ref{figRC}) for both scenarios $a>b$ and $a<b$.

We observe that for the case $a>b$, shown in Fig. \ref{figRCa}, the VdW black hole is unstable for small horizon values since both the semi-classical and {RaGr}-corrected heat capacity functions take negative values. Then, after a critical radius, the VdW black hole becomes stable. Here, we do not observe a first-order phase transition since the specific heat function is not equal to zero. In \ref{figRCb}, where $a<b$, we observe that the black hole is stable except at a highly effective {RaGr} formalism. We can conclude that if the {RaGr} effects become dominant, then the stability character of the VdW black hole can alter. 

{Finally, we would like to emphasize that this foreseen unstable-stable phase transition is also found in other scenarios \cite{OurVdW, Aydiner}.  However, depending on the particular values of the parameters, Ditta et al showed the possibility of double-phase transition for the VdW black holes \cite{Ditta2023}.}

\section{Ideal gas}\label{sec4}

The equation of state of the VdW fluid can easily be reduced to that of an ideal gas by taking $a=b=0$. In this circumstance Eq. (\ref{eq}) becomes
\begin{equation}
T=Pv, 
\end{equation}
{ and} compatible with this approach the {RaGr}-corrected lapse function, namely Eq. \eqref{RGlapse}, reduces to the following form
\begin{equation}
f\left( r\right) =\frac{8\pi P}{3}r^{2}-\frac{2M}{r}-\frac{2P\gamma \pi }{E_{P}^{2}}.
\end{equation}
Here, we note that the last term stands for the {RaGr} modification. Similarly, all other thermodynamic quantities can be rewritten for the ideal gas case. For example, the GR-corrected mass and thermodynamic volume read
\begin{eqnarray}
M&=&\frac{4}{3}\pi P r_{H}^{3}-\frac{\pi \gamma P}{E_{P}^{2}}r_{H}{-\frac{8}{27}\frac{\pi \gamma }{E_{P}^{2}}\ln 2 }, \\
V&=&\frac{4}{3}\pi r_{H}^{3}-\frac{\gamma \pi  }{E_{P}^{2}}r_{H}.
\end{eqnarray}%
{ We observe that the Ra-Gr corrections occur in negative terms in mass and volume functions. Therefore, we conclude that both functions are always smaller than their semi-classical values. This fact also implies that these functions are limited by a lower event horizon value.

In this case, the Ra-Gr Hawking temperature reads:
\begin{equation}
T_{H}=2r_{H}P\bigg(1-\frac{3\gamma }{4r_{H}^2{E_{P}^{2}}}\bigg).
\end{equation}
Similar to the previous functions, we notice that the RaGr-correction term in the Hawking temperature is negative. We find that the Hawking temperature is physical if and only if the event horizon satisfies the following condition
\begin{eqnarray}
r_{H}\geq \sqrt{\frac{3\gamma }{4E_{p}^{2}}}.
\end{eqnarray}
To illustrate this, we depict the Hawking temperature function versus the event horizon in Fig. \ref{idealgasT}.

\begin{figure}[tbh!]
\centering
\includegraphics[scale=1]{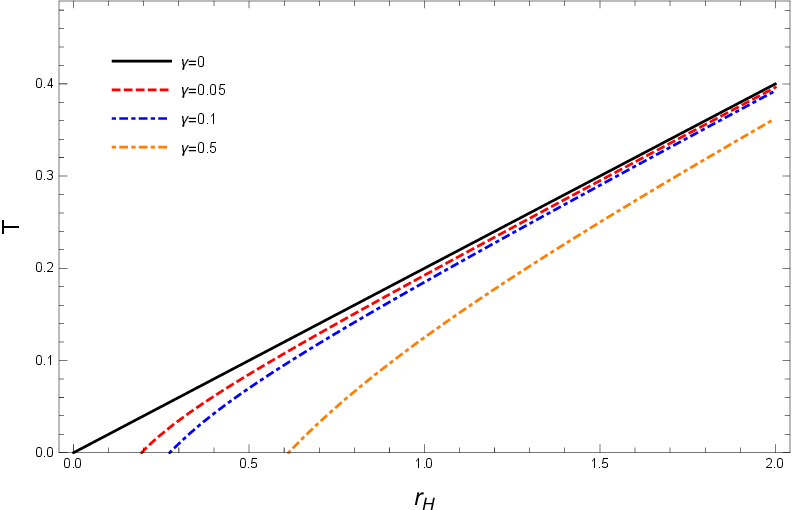}
\caption{The variation of the {RaGr}-corrected Hawking temperature of the Ideal
gas black hole versus $r_{H}$ for $P=0.1.$}
\label{idealgasT}
\end{figure}
We see that in the ordinary case, $\gamma=0$, the Hawking temperature increases monotonically and it is physical for all event horizon values. Within the context of Ra-Gr formalism, we observe that the Hawking temperature is always smaller than its original form. Here, we detect the lower bound of the horizon and we find that for greater deformation parameter values the physical event horizon interval decreases.

Next, to ascertain whether there is a phase change due to the effect of the Ra-Gr formalism, we plot the $P-r_{H}$ isotherm graph.  

\begin{figure}[tbh!]
\centering
\includegraphics[scale=1]{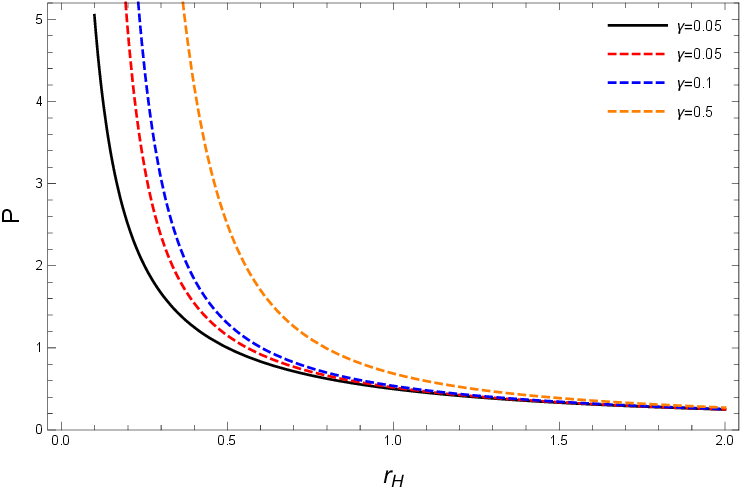}
\caption{RaGr-corrected isotherm of the Ideal gas black hole for $T_H=1.$}
\label{idealgasP}
\end{figure}
We see that with Ra-Gr corrections, the pressure takes larger values and a critical value does not emerge \cite{swallowtail}. Like in the ordinary ideal gas case, we do not observe a phase transition structure and swallowtail behavior in the Ra-Gr corrected case.

Finally, we derive the Ra-Gr corrected specific heat function
\begin{eqnarray}
C\simeq 2\pi r_{H}^{2}-\frac{2\pi \gamma }{E_{p}^{2}}.  
\end{eqnarray}
and depict it versus the event horizon in Fig. \ref{idealgasC}

\begin{figure}[tbh!]
\centering
\includegraphics[scale=1]{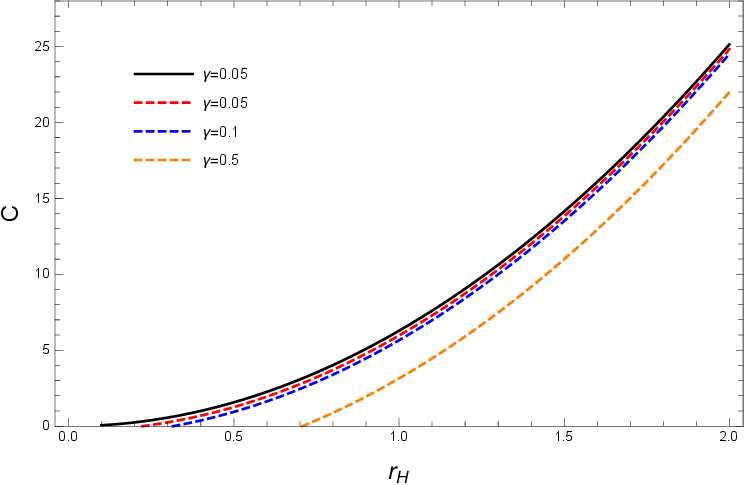}
\caption{The variation of the RaGr-corrected specific heat of the Ideal
gas black hole versus $r_{H}$.}
\label{idealgasC}
\end{figure}
We observe that in the physically meaningful region, the heat capacity is always positive, which means that the ideal gas BH is in stable equilibrium in the Ra-Gr formalism.
}

\section{Conclusion}
In this work, we investigated the effects of the rainbow gravity formalism on the black hole thermodynamics that mimics the thermodynamics of Van der Waals fluid and ideal gas. We explored the impacts by analyzing {RaGr}-corrected mass, temperature, thermodynamic pressure, entropy, and specific heat functions. We found that the mass function increases monotonically. The {RaGr} formalism does not alter their characteristics, however, it sets a lower bound value on the event horizon radius. We then found that these lower values are the minimum radii at which the black holes are physical, in other words, they are stable. Then, using the mass function we derived the thermodynamic volume function. We showed that similar to the mass function, the {RaGr} formalism decreased the volume function. Next, we examined the Hawking temperature. In a particle choice of parameter, we observed an initial decrease behavior, however, later we found that that interval is not in the physical range. Therefore, we ignored that unphysical behavior and we concluded that the VdW black hole's temperature increases monotonically. We also noticed that the {RaGr} corrections decreased the value of the Hawking temperature. Next, we derived the entropy function. We found that the entropy function altered with a positively contributing term. Then, we discussed the specific heat function. We noted that neither a first-order phase transition nor a remnant value exists. We showed the VdW black holes are stable starting from a lower bound event horizon radius. Finally, we reduced our comprehensive analysis to a subproblem. Similar to the VdW black hole thermodynamics, we wrote down the mass, thermodynamic volume, and Hawking temperature for ideal gas black holes. We found that {RaGr}-formalism also alters them similar to the VdW black hole case. 

\section*{Acknowledgments}
{
The authors thank the anonymous
referees for helpful suggestions and enlightening comments, which helped to improve the quality of this paper.} This work is supported by the Ministry of Higher Education and Scientific
Research, Algeria under the code: B00L02UN040120230003. B. C. L\"{u}tf\"{u}o%
\u{g}lu is grateful to the P\v{r}F UHK Excellence project of 2211/2023-2024
for the financial support.

\section*{Data Availability Statements}

The authors declare that the data supporting the findings of this study are
available within the article.

\section*{Competing interests}

The authors declare no competing interests.

\end{document}